\documentclass[aps,prl,twocolumn,epsfig,pdflatex]{revtex4-1}

\usepackage{amsmath,amssymb}
\usepackage{graphicx}

\usepackage{float}
\usepackage[usenames,dvipsnames]{xcolor}
\usepackage{amsthm,comment}
\usepackage{booktabs}
\usepackage[export]{adjustbox}
\usepackage[section]{placeins}
\usepackage[caption=false]{subfig}
\usepackage[percent]{overpic}
\bibpunct{[}{]}{,}{n}{}{}
\definecolor{red1}{HTML}{FF4136}
\definecolor{green1}{HTML}{00802b}

\usepackage[hidelinks]{hyperref} 

\hypersetup{colorlinks=true,citecolor=Red,linkcolor=Blue,urlcolor=Blue}
\usepackage[export]{adjustbox}

\graphicspath{{Figures/}}

\begin{document}

\title {Prediction of exotic magnetic states in the alkali metal quasi-one-dimensional \\ iron selenide compound  Na$_2$FeSe$_2$ }

\author{Bradraj Pandey$^{1,3}$, Ling-Fang Lin$^{1,2}$, Rahul Soni$^{1,3}$, Nitin Kaushal$^{1,3}$, Jacek Herbrych$^{4}$, Gonzalo Alvarez$^{5}$,  
and Elbio Dagotto$^{1,3}$}
\affiliation{$^1$Department of Physics and Astronomy, University of Tennessee, Knoxville, Tennessee 37996, USA \\ 
$^2$School of Physics, Southeast University, Nanjing 211189, China \\ 
$^3$Materials Science and Technology Division, Oak Ridge National Laboratory, Oak Ridge, Tennessee 37831, USA \\ 
$^4$Department  of  Theoretical  Physics,  Faculty  of  Fundamental  Problems  of  Technology, Wroclaw  University  of  Science  and  Technology,  
50-370  Wroclaw,  Poland\\
$^5$Computational Sciences $\&$ Engineering Division and Center for Nanophase Materials Sciences,
Oak Ridge National Laboratory,~Oak Ridge,~Tennessee 37831,~USA
 }

\begin{abstract}
The magnetic and electronic phase diagram of a model for the quasi-one-dimensional alkali metal iron selenide
compound Na$_2$FeSe$_2$ is presented. The novelty of this material is that the valence of iron is Fe$^{2+}$
contrary to most other iron-chain compounds with valence Fe$^{3+}$.
Using first-principles techniques, we developed a three-orbital tight-binding
model that reproduces the {\it ab initio} band structure near the Fermi level. Including Hubbard and Hund couplings and
studying the model via the density matrix renormalization group and Lanczos methods, we constructed the ground state phase diagram.
A robust region where the block state  $\uparrow \uparrow \downarrow \downarrow 
\uparrow \uparrow \downarrow \downarrow$ is stabilized was unveiled. The analog state in iron ladders, employing 2$\times$2 ferromagnetic blocks,
is by now well-established, but in chains a block magnetic order has not been observed yet in real materials. 
The phase diagram also contains a large region of canonical staggered spin order
$\uparrow  \downarrow \uparrow \downarrow \uparrow  \downarrow \uparrow$ at very large Hubbard repulsion.
At the block to staggered transition region, a novel phase is stabilized with a mixture of both states: an inhomogeneous
orbital-selective charge density wave with the exotic spin configuration 
$\uparrow \uparrow \downarrow  \uparrow \downarrow \downarrow \uparrow \downarrow$. 
Our predictions for Na$_2$FeSe$_2$ may guide crystal growers and neutron scattering experimentalists
towards the realization of block states in one-dimensional iron-selenide chain materials.
\end{abstract}


\maketitle

\section{I. Introduction}

Iron-based pnictides and selenides are fascinating materials with exotic magnetic and superconducting properties~\cite{fernandes,pdai,elbio}.
For iron-selenides the low-temperature insulating ground state has robust local magnetic moments~\cite{chuang,gretar,Bao},
highlighting the importance of Hubbard and Hund coupling interactions among the electrons occupying the $3d$ orbitals~\cite{pdai,elbio}. 
The competition between charge, spin, lattice, and orbital degrees of freedom can give rise to various types of exotic magnetic 
and electronic ordering. In particular, recently 
the  two-leg ladder iron selenide materials have received considerable attention. One reason
is their similarity with copper-based ladders, with a spin gap in the undoped
limit and superconductivity upon doping by high pressure~\cite{riera,rice}.
Moreover, in the two-leg ladder iron-based compound BaFe$_2$Se$_3$, an exotic block-antiferromagnetic (AFM) order
(involving $ 2 \times 2 $ ferromagnetically aligned blocks, coupled antiferromagnetically along the legs of the ladder)
has been reported using inelastic neutron diffraction methods~\cite{mourigal,caron,caron2,wang2016,lei,mouri}, confirming earlier predictions by theory~\cite{julian,julian2}. 
BaFe$_2$Se$_3$ is an insulator with robust N\'{e}el temperature $T_N \sim$ 250K into the block phase and 
large individual magnetic moments $\sim 2.8 \mu_B$. In another  iron-based ladder material, 
where  K replaces Ba  leading to KFe$_2$Se$_3$, the magnetic moments align ferromagnetically along the rungs 
but antiferromagnetically along the legs forming 2$\times$1 blocks~\cite{caron}.

In addition to these ladder materials, there are some experimentally observed iron-selenide compounds, 
such as TlFeS$_2$, TlFeSe$_2$ and KFeSe$_2$, which contain weakly coupled quasi-one-dimensional {\it chains}~\cite{bronger,seidov}.
In these compounds iron is in a valence Fe$^{3+}$, corresponding to $n=5$ electrons in the $3d$ iron orbitals.
 Based on magnetic susceptibility, electric resistivity, and electron-spin resonance, 
TlFeSe$_2$ behaves as a quasi one-dimensional standard spin-staggered antiferromagnet~\cite{veliyev}. 
Furthermore, neutron diffraction experiments on TlFeS$_2$ also indicate~\cite{nishi} staggered spin order below $T_N=196$K.

\begin{figure}[h]
\centering
\rotatebox{0}{\includegraphics*[width=\linewidth]{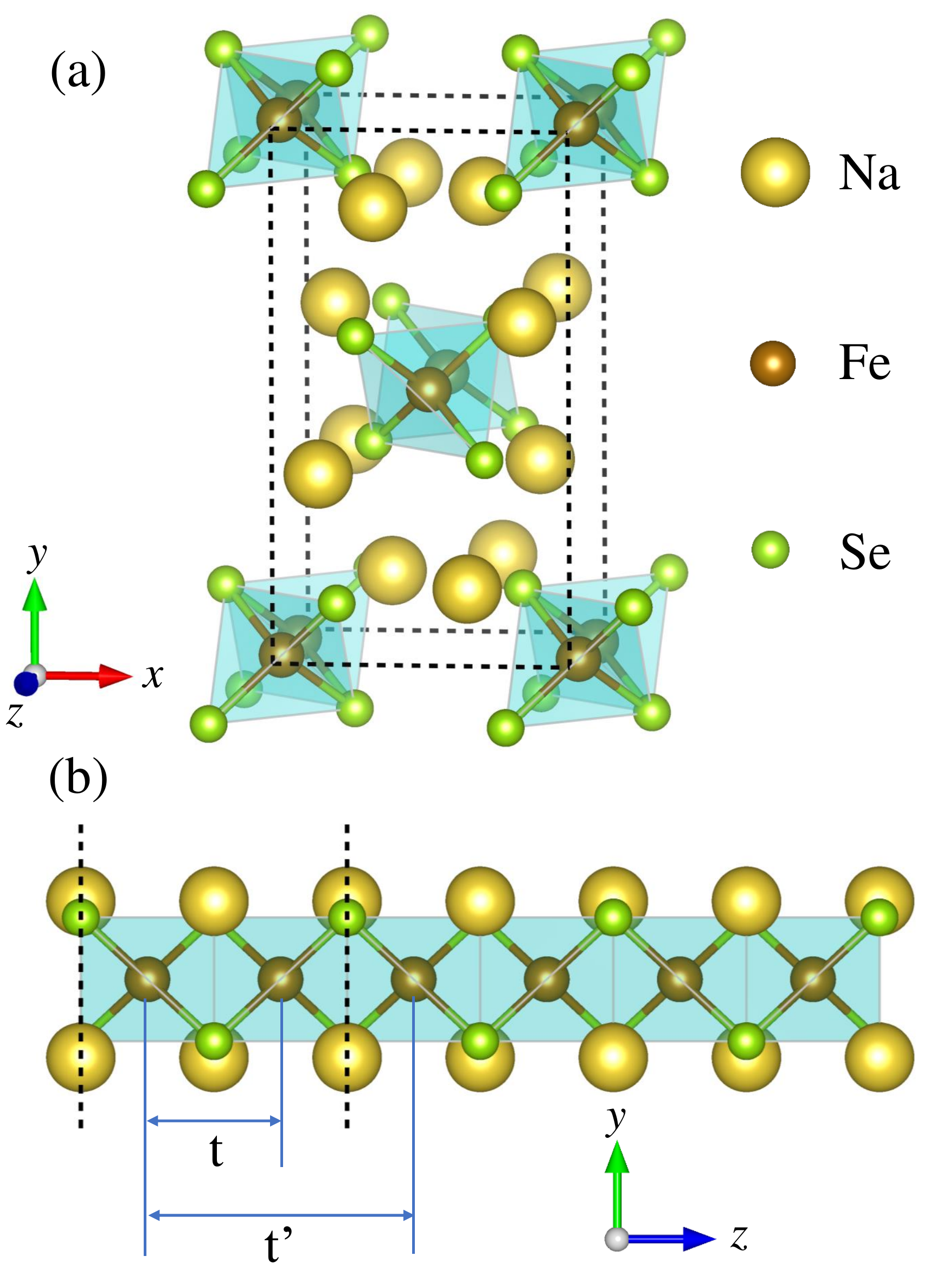}}
\caption{(a) Crystal structure of  Na$_2$FeSe$_2$, the material that we predict should present exotic magnetic order. 
(b) Side view of a single Fe chain and the nearest-neighbors 
$t$ and next-nearest-neighbors $t'$ hopping amplitudes used in our study.}
\label{fig1}
\end{figure}
The experimental developments described above in quasi-1D iron-based materials provides a playground
 for theoretical many-body calculations based on multi-orbital Hubbard model~\cite{nocera,ORVB,herbrychNatComm,herbrychblock,patelComm}. Using accurate 
numerical techniques for low-dimensional systems, such as the density matrix renormalization group method (DMRG)~\cite{white1,steven}, 
the high-pressure superconducting two-leg ladder compound BaFe$_2$S$_3$~\cite{takahashi,yamauchi,petrovic} was explored
with regards to magnetic and pairing properties keeping two orbitals active~\cite{patel,arita}. Evidence for the correct rung-FM 
and leg-AFM spin order was found over large portions of interaction parameters~\cite{patel}.
Evidence of metallization under high pressure was also reported~\cite{yangpressure1,yangpressure2}. This is considered 
a precursor of superconductivity, which was also shown to appear in theoretical studies of 
two-orbital one-dimensional models upon hole doping~\cite{nocera,ORVB}.
Even multiferroicity was unveiled in iron ladders~\cite{dongmultiferroic}, indicating an unexpected rich behavior. Moreover,
novel Te-based ladders were predicted to display interesting magnetic properties as well~\cite{yang1,yang2}.

The phase diagram of a three-orbital Hubbbard model for chains, 
was also studied using DMRG~\cite{julian,julian2}, unveiling various types of exotic magnetic and electronic phases. 
More canonical
ferromagnetic and staggered $\uparrow \downarrow \uparrow \downarrow$ states were also stabilized
varying the Hund and Hubbard interaction parameters. The  spin dynamical properties of exotic  orbital-selective Mott phases
(displaying the selective localization of electrons on a particular orbital) were also analyzed, revealing unusual coexisting modes of spin excitations~\cite{herbrychNatComm}.

The magnetic phase diagram of the five-orbital Hubbard model for iron-selenide materials was initially studied 
using real-space Hatree-Fock approximations for chains~\cite{german} and ladders~\cite{luo}. At electronic density $n=5$, 
relevant to previously known chain compounds such as TlFeSe$_2$, a simple staggered AFM phase in a
large parameter space of the phase diagram was reported, in agreement with existing experiments. 
Interestingly, a much richer phase diagram was theoretically predicted
for chains with the electronic density $n=6$. More reliable DMRG studies of the three-orbital Hubbard model at $n=6$ have also 
consistently reported a similar wide variety of exotic phases for $n=6$, including the block phase $\uparrow \uparrow \downarrow \downarrow$ at robust
Hund coupling~\cite{julian}, as well as generalizations to longer blocks~\cite{herbrychblock} and even spontaneously 
formed spiral phases~\cite{herbrychspiral}. But thus far only two-leg
ladder materials, such as BaFe$_2$Se$_3$ and BaFe$_2$S$_3$, have been studied experimentally, confirming 
the block nature of the spin state -- either 2$\times$2 or 2$\times$1 blocks -- as well as exotic
superconductivity upon high pressure. However, finding a truly $n=6$ one-dimensional version, with only chains instead of ladders, would
add another interesting member to the existing group of realizations of the theory predictions, opening a novel avenue for research.

Recently, the possibility of preparing the alkali iron selenide compound Na$_2$FeSe$_2$ has been discussed~\cite{stuble}.
In  Na$_2$FeSe$_2$ the iron atom is in a valence state Fe$^{2+}$, which correspond to an electronic density
$n=6$ for the $3d$ Fe orbitals. As already discussed, Hartree-Fock studies of low-dimensional multiorbital models 
with electronic density $n=6$ displayed a much richer phase diagram with exotic phases, as compared to the
canonical staggered order of the $n=5$ case. Motivated by the recent experimental efforts~\cite{stuble}, 
in this publication we study theoretically the magnetic and electronic properties of the chain compound  Na$_2$FeSe$_2$. 
Using first principles calculations we obtain the relevant hopping amplitudes. 
Next, using computationally accurate techniques, such as DMRG and Lanczos methods, 
we construct the ground-state phase diagram by varying the on-site same-orbital Hubbard $U$ repulsion and 
the on-site Hund coupling $J_H$. At low values of $J_H/U$, the staggered AFM order with wavevector $\pi$ dominates 
in a large portion of the phase diagram. However, increasing $J_H/U$ into the realistic regime for iron-based compounds, 
interesting block phases, particularly $\uparrow \uparrow \downarrow \downarrow$,  
dominate in a large region of parameter space. In contrast to Hartree-Fock methods, DMRG and Lanczos take into account 
quantum fluctuations rendering the results more reliable. Finally, albeit in a narrow region of parameter space, a novel phase
$\uparrow \uparrow \downarrow \uparrow \downarrow \downarrow \uparrow \downarrow$ was also found with a mixture of 
properties of the dominant block and staggered states.

The organization of the paper is as follow. In Section II, details of the {\it ab initio} calculations are described. 
Section III contains the three-orbital Hubbard model and details of the numerical methods. Section IV presents
the DMRG and Lanczos predictions, where first we focus on the results at the realistic Hund coupling $J_H/U=1/4$, 
and later an extended phase diagram of the model is provided. Finally, conclusions are provided in Section V.

\section{II. Ab Initio Calculations}

\begin{figure}[h]
\centering
\rotatebox{0}{\includegraphics*[width=\linewidth]{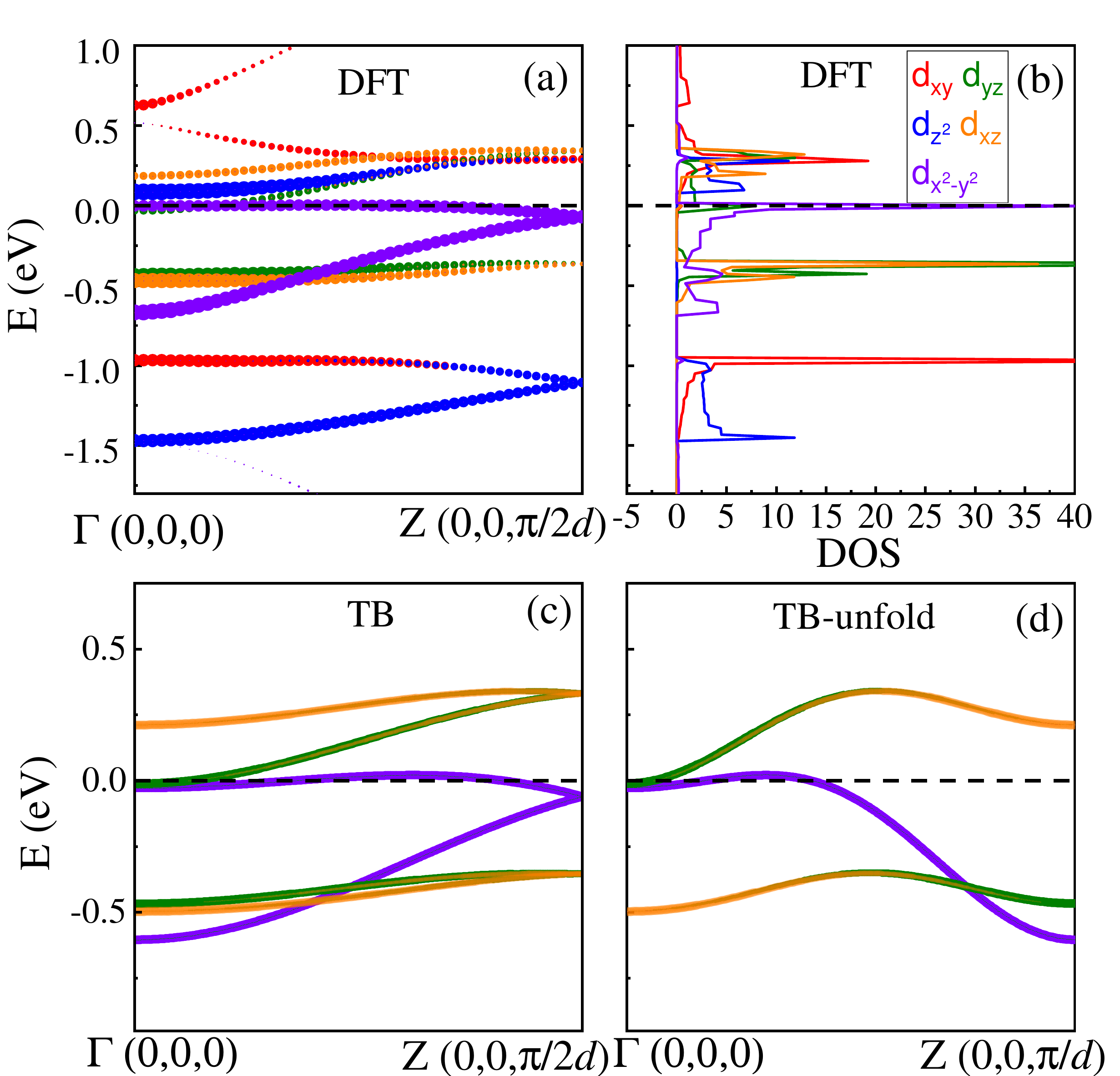}}
\caption{(a) Band structure and (b) projected density-of-states of the Na$_2$FeSe$_2$ single-chain compound obtained using DFT calculations. (c)  Tight-binding (TB) band structure in the folded zone. (d)  Tight-binding unfolded (TB-unfold) band structure used in the DMRG calculations. The zero in the vertical axis is the position of the Fermi level.}
\label{fig2}
\end{figure}

The crystal structure of Na$_2$FeSe$_2$ is shown in Fig.~\ref{fig1}.
The most prominent feature is that edge-sharing 
FeSe$_2$ tetrahedral form one-dimensional chains running along the $c$ axis. 
Here, first-principles density functional theory (DFT) calculations are used employing 
the lattice constants $a$, $b$, and $c$, and the atomic positions of the Na, Fe, and Se atoms as 
reported in Ref.~\onlinecite{stuble}. These lattice constants are $a=6.608 {\rm \AA}$, $b=11.903 {\rm \AA}$, and $c=5.856 {\rm \AA}$. 
The space group is Ibam (no. 72), and the atomic positions of Na(8j), Fe(4a), and Se(8j) are 
(0.1562, 0.35565, 0.0), (0.0, 0.0,0.25), and (0.21638, 0.11435, 0.0), respectively. 
The band structure and the projected density-of-states for the 3$d$ orbitals are presented in Fig.~\ref{fig2}(a,b).
The orbital $d_{x^2-y^2}$ contributes primarily near the Fermi level. 
The contribution of the orbitals $d_{xz}$ and $d_{yz}$ is subdominant, but not negligible.

Considering the one-dimensional character of the atomic structure, all interchain electron hopping amplitudes are neglected
and we only focus on the intrachain hoppings. In other words, 
only a Na$_2$FeSe$_2$ single chain [shown in Fig.~\ref{fig1}(b)] is considered in the DFT procedure. 
The calculations were performed using the generalized gradient approximation~\cite{perdew} and 
the projector augmented wave (PAW) pseudopotentials~\cite{blochl}, implemented in the Vienna 
{\it ab initio} Simulation Package (VASP) code~\cite{kresse,joubert}.
Since the magnetic properties will be considered via many-body calculations,
magnetism was not included in the derivation 
of the bands and hopping amplitudes from first principles. Following a self-consistent calculation with total energy 
convergence of order eV, the maximally localized Wannier functions~\cite{marzari} were constructed using the WANNIER90 
code~\cite{mosto} from the {\it ab initio} ground-state wave function. 

We constructed three Wannier functions involving the orbital basis {$d_{xz}$, $d_{yz}$, $d_{x^2-y^2}$} 
for each iron and deduced the hopping parameters, readjusted to fit properly the band structure after reducing the original
five orbitals to three (see Sec.~III for details). The corresponding band structure 
using these hoppings is displayed in Fig.~\ref{fig2}(c), which agrees well with the DFT band structure.
Note that there are two Fe atoms in the primitive unit cell in the DFT calculation because of the alternating
positions of the Se atoms, leading to a unit cell of $2d$ length, where $d$ is the distance between two nearest-neighbor iron atoms. 
The band structure can be unfolded since we only focus on the irons only, leading to the bands in Fig.~\ref{fig2}(d) that 
were used in the DMRG calculations.

\section{III. Model and Method}

The Hamiltonian for  the one-dimensional chain of Na$_2$FeSe$_2$, with three orbitals at each iron 
site, will be described by the multi-orbital Hubbard $H=H_k+H_{in}$. 
The kinetic or tight-binding component contains the nearest- and next-nearest neighbors hopping:
\begin{eqnarray}
H_k = \sum_{i,\sigma,\gamma,\gamma'} t_{\gamma, \gamma'}\left(c^{\dagger}_{i\sigma,\gamma}c^{\phantom\dagger}_{i+1, \sigma, \gamma'}+H.c.\right) \nonumber \\
+  t'_{\gamma, \gamma'}\left(c^{\dagger}_{i\sigma,\gamma}c^{\phantom\dagger}_{i+2, \sigma, \gamma'}+H.c.\right)+ \sum_{i\gamma \sigma} \Delta_{\gamma} n_{i,\sigma \gamma},
\end{eqnarray}
where $t_{\gamma, \gamma'}$ is the nearest-neighbor (NN)  3$\times$3 
hopping amplitude matrix between sites $i$ and $i+1$ in the orbital space $\gamma=\{d_{xz},d_{yz},d_{x^2-y^2}\}$.
$n_{i,\sigma \gamma}$ stands for the orbital and spin resolved particle number operator.
These orbitals will be referred to as $\gamma = \{1,2,3\}$, respectively, in the remaining of the paper, 
for notation simplicity. As explained before, the hopping matrices for Na$_2$FeSe$_2$ were obtained from 
a tight-binding Wannier function analysis of first-principles results and they are in eV units. 
Explicitly, the NN 3$\times$3 matrix $t_{\gamma, \gamma'}$ between sites $i$ and $i+1$ in orbital space is given by:
\[
t_{\gamma, \gamma'}=
  \begin{bmatrix}
    -0.177 & 0.171 & 0.000 \\
    -0.171 & 0.114 & 0.000  \\
     0.000 & 0.000 & 0.144 
  \end{bmatrix}
\] 
where $\gamma$ are the orbitals for site $i$ and  $\gamma'$ for site $i+1$.
$t'_{\gamma, \gamma'}$ is the NNN hopping matrix between sites $i$ and $i+2$:
\[
t'_{\gamma, \gamma'}=
  \begin{bmatrix}
    -0.037 & -0.003 & 0.000 \\
     0.003 & -0.053 & 0.000  \\
     0.000 &  0.000 & -0.064 
  \end{bmatrix}
\]
The on-site matrix containing the crystal fields $\Delta_{\gamma}$ for each orbital is given by:
\[
t^{OnSite}_{\gamma, \gamma}=
  \begin{bmatrix}
    -0.068 & 0.000 & 0.000 \\
     0.000 & -0.134 & 0.000  \\
     0.000 &  0.000 & -0.188 
  \end{bmatrix}
\]
 
Note that we follow the convention that  each $3 \times 3$ matrix (both $t_{\gamma, \gamma'}$ and $ t'_{\gamma, \gamma'}$) represent 
the hopping matrix to move from one iron site to another. The full hopping  matrix, which includes both the 
back and forth hopping processes, are of size $6 \times 6$ containing $t_{\gamma, \gamma'}$ in the upper off-diagonal 
block, the transpose of  $t_{\gamma, \gamma'}$ in the lower
off-diagonal block, and the on-site matrix $t^{OnSite}_{\gamma,\gamma}$ in both diagonal blocks~\cite{luo}.
The kinetic energy bandwidth is $W=0.94$~eV.

The electronic interactions portion of the Hamiltonian is standard:
\begin{eqnarray}
H_{in}= U\sum_{i\gamma}n_{i\uparrow \gamma} n_{i\downarrow \gamma} +\left(U'-\frac{J_H}{2}\right) \sum_{i,\gamma < \gamma'} n_{i \gamma} n_{i\gamma'} \nonumber \\
-2J_H  \sum_{i,\gamma < \gamma'} {{\bf S}_{i,\gamma}}\cdot{{\bf S}_{i,\gamma'}}+J_H  \sum_{i,\gamma < \gamma'} \left(P^+_{i\gamma} P_{i\gamma'}+H.c.\right). 
\end{eqnarray}
The first term is the Hubbard repulsion between electrons in the same orbital. 
The second term is the electronic repulsion between electrons at 
different orbitals where the standard relation $U'=U-2J_H$ is assumed. The third term represents the Hund's 
interaction between electrons occupying the active $3d$ orbitals. 
The operator ${\bf S}_{i,\gamma}$ is the total spin at site $i$ and orbital $\gamma$. The fourth term is the
pair-hopping between different orbitals at the same site $i$, where $P_{i, \gamma}$=$c_{i \downarrow \gamma} c_{i \uparrow \gamma}$.

To solve numerically this Hamiltonian and obtain the ground state properties of Na$_2$FeSe$_2$, 
the DMRG and Lanczos methods were used. Open boundary conditions were employed in DMRG and at least 1200 
states kept during the calculations. For these DMRG calculation, we used the DMRG++ computer program~\cite{gonzalo}. 
We fixed the electronic density per-orbital to be $n=4/3$ (four electrons per site, i.e.
four electrons in three orbitals). 
Such electronic density is used in the context of iron superconductors where 
iron is in a valence Fe$^{2+}$, corresponding to 6 electrons in five orbitals. A common simplification is to drop
one orbital doubly occupied and one empty, leading to 4 electrons in the remaining three orbitals.
Most of the DMRG calculations were performed using chains of length $L=16$ and $L=24$ which for our purposes of
finding the magnetic properties of the ground state are sufficient. 
Furthermore, by investigating small lattice sizes ($L=4$) with exact Lanczos diagonalization we reached the 
same conclusions.

\section{IV. Results}
\begin{figure}[h]
\centering
\rotatebox{0}{\includegraphics*[width=\linewidth]{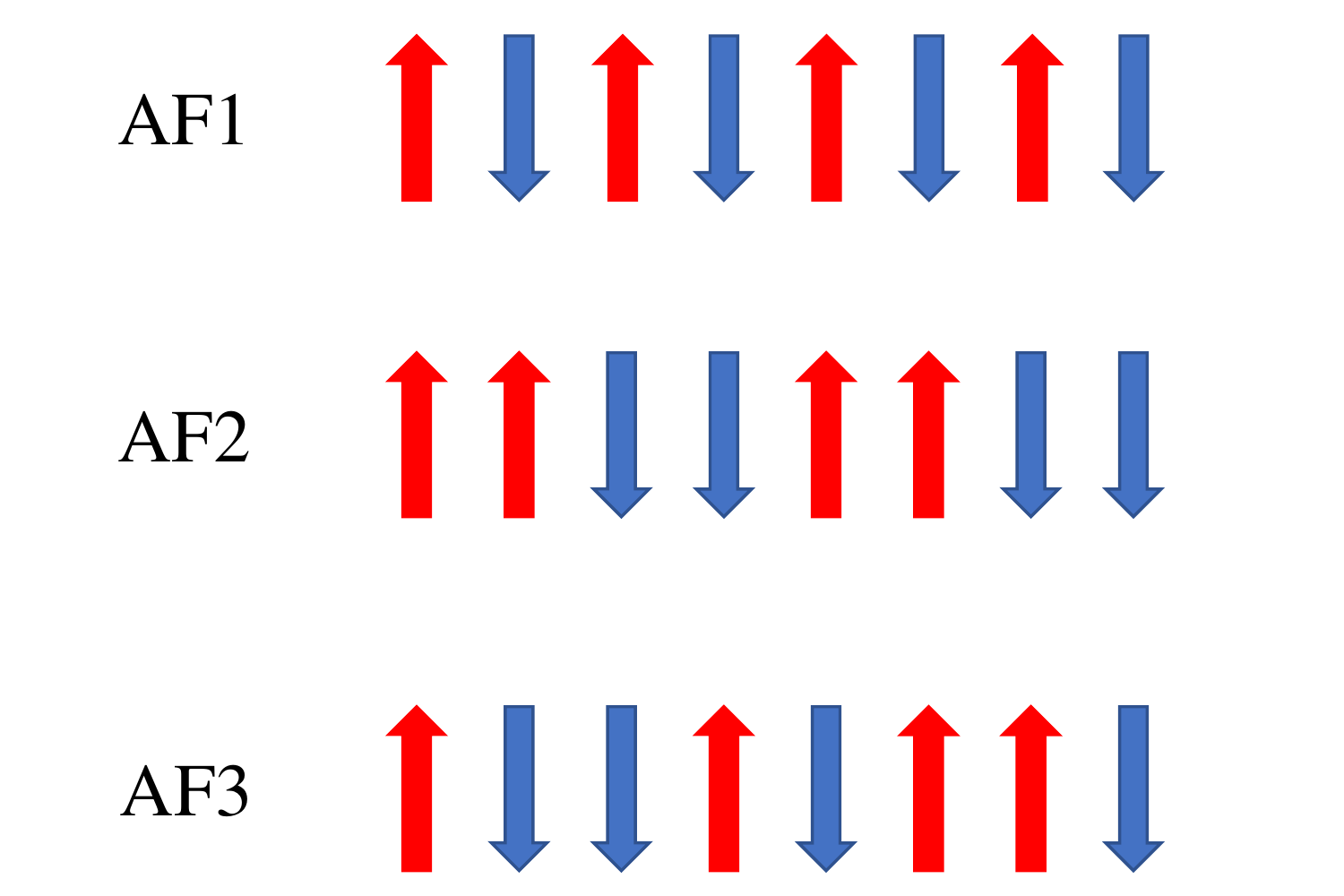}}
\rotatebox{0}{\includegraphics*[width=\linewidth]{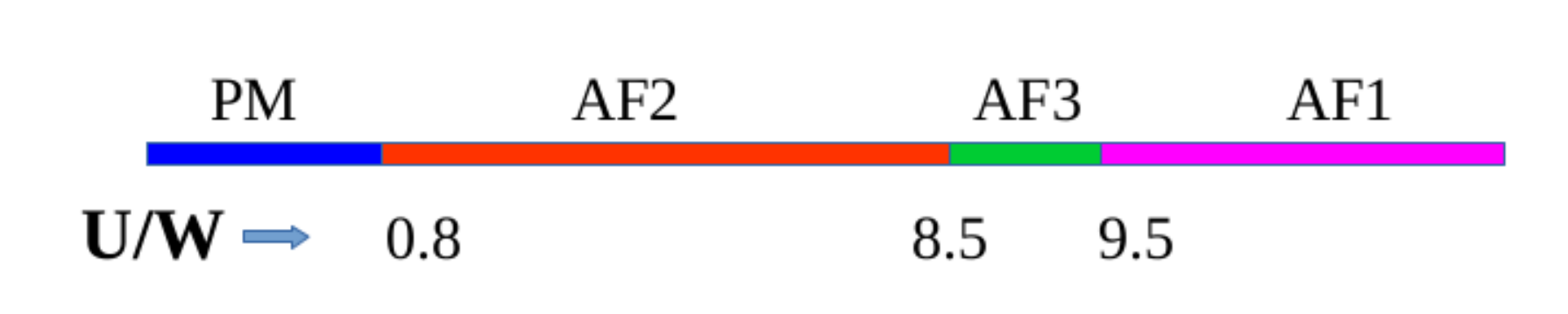}}
\caption{Schematic representation of the magnetic states observed in the phase diagram. (i) AF1:
 standard staggered antiferromagnetic phase $\uparrow \downarrow \uparrow \downarrow$;
(ii) AF2: antiferromagnetically-coupled ferromagnetic blocks resulting in $\uparrow \uparrow \downarrow \downarrow$ spin-order. 
(iii) AF3: mixed ferro and antiferro magnetic ordering $\uparrow \uparrow \downarrow \uparrow \downarrow \downarrow \uparrow \downarrow$
 stable in a narrow region of couplings.
At the bottom: schematic phase diagram of the ground state for fixed $J_H/U=0.25$.
} 
\label{fig3}
\end{figure}
\begin{figure}[h]
\centering
\rotatebox{0}{\includegraphics*[width=\linewidth]{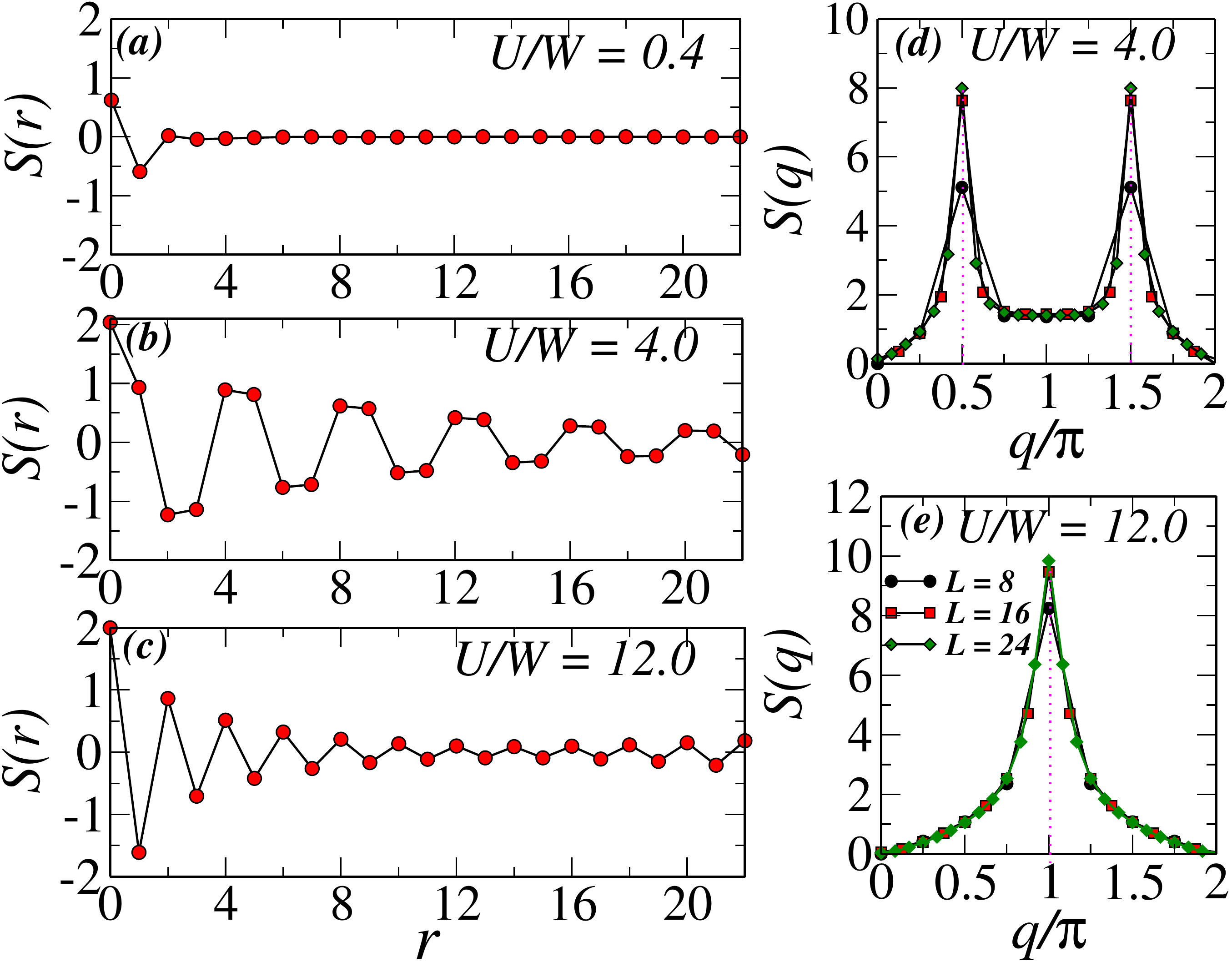}}
\caption{Real-space spin correlation $S(r)=\langle {\bf S}_m \cdot {\bf S}_l \rangle$, with $r=|m-l|$, 
for various values of the Hubbard interaction $U/W$, at fixed $J_H/U=0.25$ and using a $L=24$ cluster studied with DMRG. 
Results are shown for (a) the PM phase at  $U/W = 0.4$, (b) the block phase (AF2) at $U/W = 4.0$, and (c)
the staggered AF1 phase at $U/W = 12.0$. The AF3 state will be discussed later in Fig.~\ref{fig6}.
 The spin structure factor $S(q)$ is shown for three values of $L$ at (d) $U/W=4.0$ in the block AF2 phase
 and (e) $U/W=12.0$ in the AF1 phase.} 
\label{fig5}
\end{figure}
In Fig.~\ref{fig3}, we show the phase diagram of the three-orbitals Hubbard model. We use realistic {\it ab initio} hopping amplitudes
for Na$_2$FeSe$_2$ and vary  $U/W$ at fixed Hund coupling $J_H/U=1/4$~\cite{luo2010}.
This phase diagram was constructed based on DMRG calculations measuring several observables:  
the site average electronic density at each orbital 
$n_{\gamma}= \frac{1}{L}\sum_{i,\sigma} \langle n_{i\sigma \gamma} \rangle$,
 the spin-spin correlation $S(r)=\langle {\bf S}_m \cdot {\bf S}_l \rangle$ (where $r=|m-l|$; $m$ and $l$ 
are sites), 
and the spin structure factor $S(q)=\frac{1}{L}\sum_{m,l}e^{-iq(m-l)}\langle  {\bf S}_m \cdot {\bf S}_l \rangle$ 
using primarily a system size $L=16$. The global electronic density is $n=$4/3 (4 electrons in three orbitals
at each site in average).

Four different phases were found: (i) a paramagnetic phase (PM) at small $U/W$, followed by
(ii) an unexpected block phase (AF2) where ferromagnetic clusters of two spins are coupled antiferromagnetically
in a $\uparrow \uparrow \downarrow \downarrow$ pattern. Then (iii) an intermediate electronically inhomogeneous 
and spin exotic state (AF3) was found, with  ferro and antiferro magnetic ordering 
$\uparrow \uparrow \downarrow \uparrow \downarrow \downarrow \uparrow \downarrow$. Finally, (iv) 
a canonical staggered antiferromagnetic phase (AF1) $\uparrow \downarrow \uparrow \downarrow$ becomes stable.
To distinguish among these magnetic phases and to obtain the approximate phase boundary location, 
we studied $S(q_{p})$ vs $U/W$, where $q=q_{p}$ is defined as the wavevector that displays a
sharp peak for each value of $U/W$ studied.   

\subsection{Results at  Hund coupling $J_H/U=1/4$}

\subsection{(a) AF2 and AF1 phases}

\begin{figure}[h]
\centering
\rotatebox{0}{\includegraphics*[width=\linewidth]{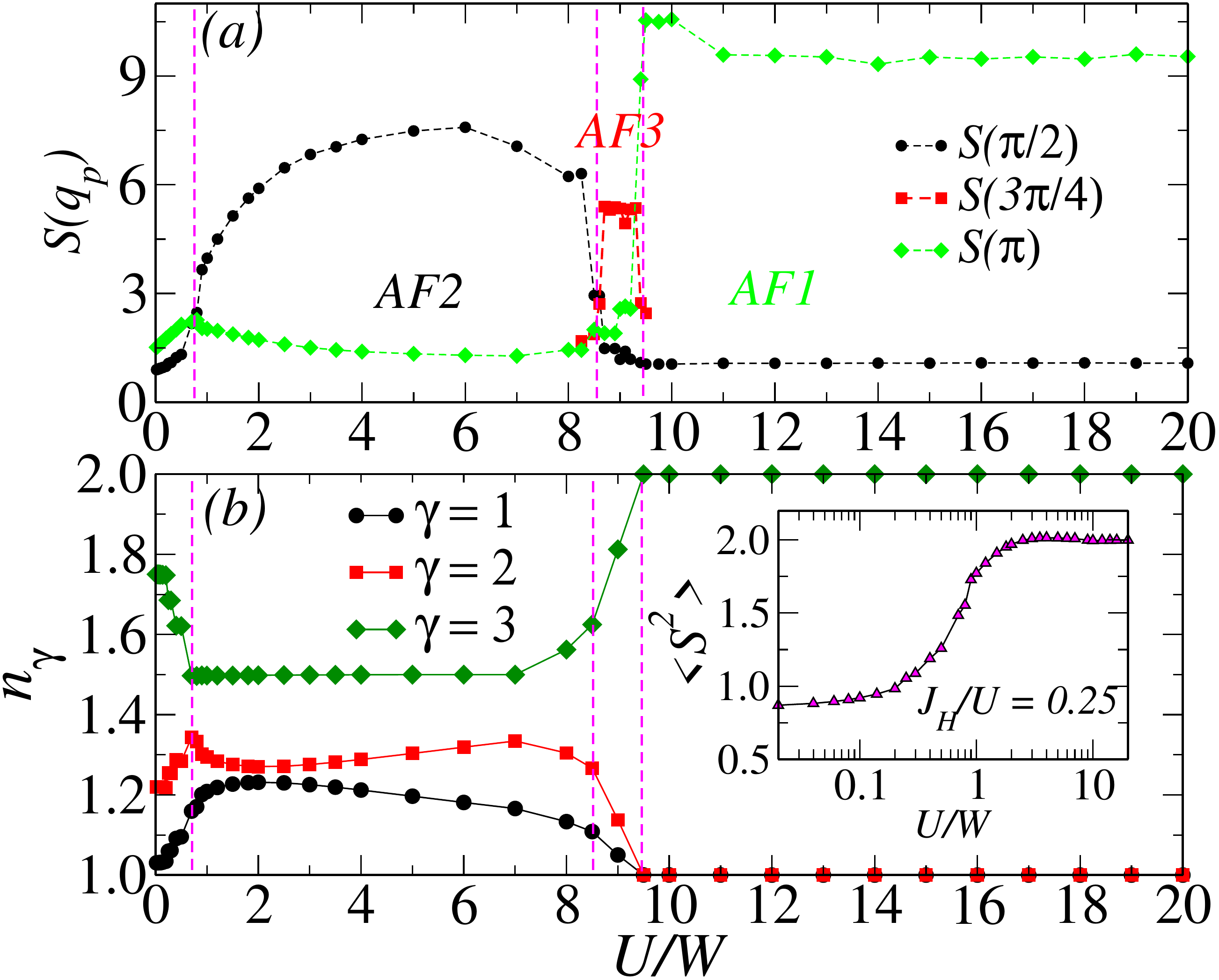}}
\caption{(a) Spin structure factor $S(q_p)$ vs $U/W$ at $J_H/U=0.25$ for several values of the three dominant
wavevectors shown in the legend.
(b) Site-average electronic occupancy $ n_{\gamma}$ for the three orbitals 
$\{ \gamma =1,2,3 \}$
vs. $U/W$ using DMRG and a chain of $L=16$ sites. {\it Inset:} site-average expectation value 
of the total spin squared vs $U/W$, at $J_H/U=0.25$.}
\label{fig4}
\end{figure}
At small Hubbard interaction $U/W$ the system displays metallic behavior without any dominant magnetic order, as expected. 
In this PM regime, the spin correlation $S(r)$ decays rapidly with distance in the range $U/W<0.8$,
as exemplified in Fig.~\ref{fig5}(a). Increasing the Hubbard interaction $U/W$, the system enters into the 
block phase with AF2 magnetic ordering.
In Fig.~\ref{fig5}(b), the spin-correlations $S(r)$ at $U/W=4.0$ are presented,
clearly showing the formation of antiferromagnetically coupled ferromagnetic spins clusters in a  
$\uparrow \uparrow \downarrow \downarrow$ pattern.
Because of  this block order, the spin structure factor $S(q)$ in the AF2 phase displays a sharp peak at $q=\pi/2$, shown in Fig.~\ref{fig5}(d). 
The peak value increases with the system size $L$ providing evidence of a stable exotic $\pi/2$-block 
magnetic state in the system. Note that the canonical power-law decaying real-space correlations in one dimension
prevents $S(q)$ from diverging with increasing $L$, but in a real material it is expected that weak interchain couplings will 
stabilize the several phases we have observed. 

As shown in Fig.~\ref{fig4}(a), $S(q_p)=S(\pi/2)$ dominates  in the range $0.8 \lesssim U/W \lesssim 8.5$, 
signalling a stable block-phase in a broad region of parameter space, at $J_H/U=0.25$. 
Similar block-AF2 spin patterns, albeit extended in two dimensions into 2$\times$2 ferromagnetic blocks, 
have been also experimentally observed in two dimensional iron-selenium based compounds with vacancies, 
such as Rb$_{0.89}$Fe$_{1.58}$Se$_2$ and K$_{0.8}$Fe$_{1.6}$Se$_2$~\cite{Bao} and more importantly for our
purposes also in the two-leg ladder BaFe$_2$Se$_3$~\cite{mourigal} which is a close ``relative'' of the
 Na$_2$FeSe$_2$ compound due to the common one-dimensionality and iron valence Fe$^{2+}$. Although it is
difficult to establish with clarity what induces this block state, previous work~\cite{julian} suggests that this phase 
is a result of competition between the Hund coupling $J_H$, favoring
ferromagnetic alignment of spins as in double-exchange manganites~\cite{dagotto2001}, and the standard superexchange 
Hubbard spin-spin interaction that  aligns the spins antiferromagnetically.
One surprising aspect is that in the block-AF2 phase the population of orbital $\gamma=3$ 
appears locked to $1.5$ in all the range of $U/W$ investigated [Fig.~\ref{fig4}(b)]. 
On the other hand, the occupancies of the other orbitals $\gamma=1$ and $\gamma=2$ 
change with varying $U/W$ in the same range.

In the inset of Fig.~\ref{fig4}(b), the mean value of the local spin-squared averaged over all sites 
 $\langle S^2 \rangle= \frac{1}{L}\sum_i \langle {\bf S}_i \cdot {\bf S}_i\rangle $ is shown vs $U/W$. 
For  $U/W>1.0$, strong local magnetic moments are fully developed 
at every  site with spin magnitude $S \approx 1$, as expected for four electrons
in three orbitals and a robust Hund coupling. In experiments, alkali metal iron selenide compounds generally
show large magnetic moments, particularly when compared to iron pnictide compounds. 

 In Fig.~\ref{fig4}(b), the site average occupancy of orbitals $ n_{\gamma}$ vs. $U/W$ is shown, and
for $U/W>9.5$ the  population of orbital $\gamma=3$ reaches 2, thus decoupling from the system, while the  
other two orbitals $\gamma = 1,2$ reach population 1. This arrangement minimizes the double occupancy at large $U/W$.
In this Mott AF1 phase, the spin correlations show a canonical staggered AFM ordering, see Fig.~\ref{fig5}(c), due to 
the dominating effect of the superexchange mechanism in the system, now involving only two active orbitals. 
The structure factor displays a sharp peak at $q=\pi$, see Fig.~\ref{fig5}(e).

\subsection{(b) Inhomogeneous AF3 phase}
\begin{figure}[h]
\centering
\rotatebox{0}{\includegraphics*[width=\linewidth]{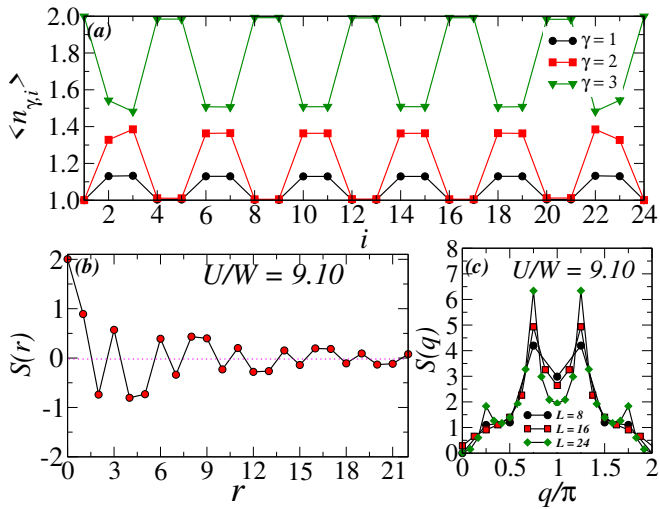}}
\caption{(a) Electronic occupancy $\langle n_{\gamma,i} \rangle$ for the three orbitals $\{ \gamma =1,2,3 \}$
vs site index $i$, at $U/W=9.1$, $L=24$, in the AF3 regime showing an orbital-selective charge density wave.
(b) Spin correlation $S(r)=\langle {\bf S}_m \cdot {\bf S}_l \rangle$ at $U/W=9.1$ using a $L=24$ cluster, 
displaying the AF3 magnetic ordering $\uparrow \uparrow \downarrow \uparrow \downarrow \downarrow \uparrow \downarrow$.
(c)  The spin structure factor for three different values of $L=8,16,24$ and at $U/W=9.1$. Clear peaks at $q=3\pi/4$ are shown.} 
\label{fig6}
\end{figure}

At interaction $8.5 <U/W< 9.5$, a novel orbital-selective charge density wave phase 
was observed, with an exotic AF3 spin ordering. This phase exists for all the lattice sizes analyzed, and
moreover it appears both using DMRG and Lanczos, as shown below, thus we believe it is a real regime of the present 
model. Figure~\ref{fig6}(a) displays the population of the three orbitals $\langle n_{\gamma,i} \rangle$ 
vs. the site index $i$, at $U/W=9.1$. The results show an orbital-selective charge density wave phase. 
The pattern that develops has two sites with integer fillings, such as 1.0 and 2.0, followed by two sites 
with a fractional filling for all the three orbitals. Orbital 3 jumps from population 2.0 as in the phase AF1,
 to population 1.5, as in the phase AF2, as compared with  Fig.~\ref{fig4}(b). The other two orbitals 1 and
2 display similar characteristics, namely a mixture of AF1 and AF2 features.

Interestingly, in parallel to an inhomogeneous charge density arrangement, a novel spin pattern AF3 
$\uparrow \uparrow \downarrow \uparrow \downarrow \downarrow \uparrow \downarrow$,
develops in the system for this range of $U/W$, see Fig.~\ref{fig6}(b).
The structure factor $S(q)$ shows a peak at $q=3\pi/4$, which grows with increasing the system size, see Fig.~\ref{fig6}(c).  
The phase boundary of this exotic AF3 phase is determined by comparing the peaks of the spin structure factors.
As shown in Fig.~\ref{fig4}(b), the peak at $q=3\pi/4$ clearly dominates over other peaks of $S(q)$ in the range 
$8.5 <U/W<9.4$.  Similar spin configurations  have also been reported in the study of the one-dimensional two-orbital 
Hubbard model~\cite{herbrychblock} at density $n=2.33$. 
We believe that this exotic phase stabilizes in the phase diagram mainly 
due to the NNN hopping $t'_{\gamma,\gamma'}$ since it generates frustration
in the system. Eventually, for large enough values of the Hubbard interaction $U/W>9.5$, the 
system enters into the insulating Mott phase with staggered AF1 magnetic ordering.

\subsection{(c) Density of states and charge fluctuations}

\begin{figure}[h]
\centering
\rotatebox{0}{\includegraphics*[width=\linewidth]{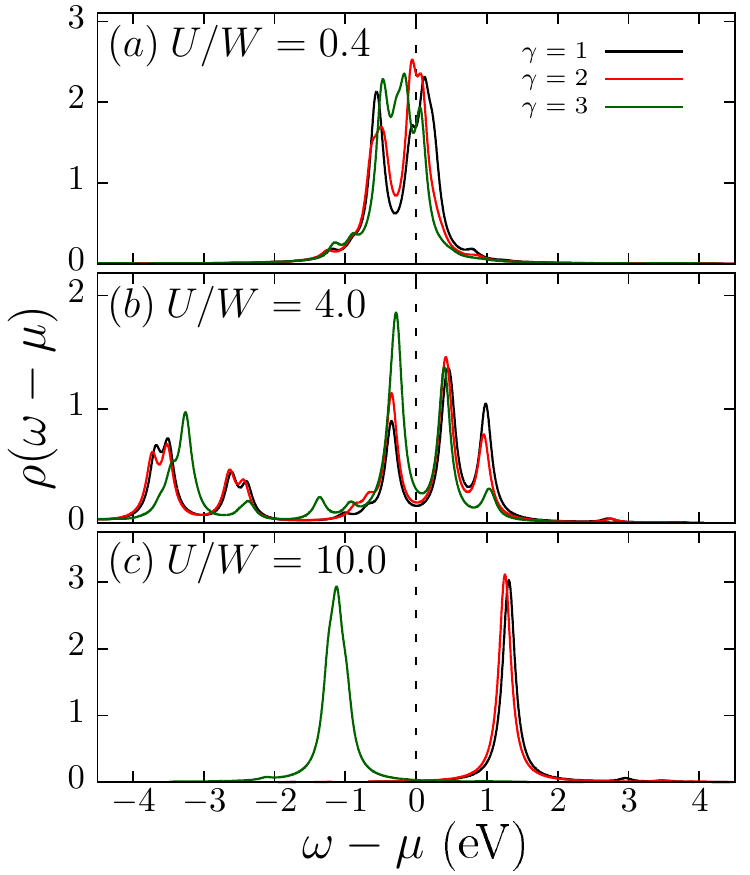}}
\caption{DOS of different orbitals corresponding to different phases at $J_H/U = 0.25$ on a 4-site 
three-orbital system using Lanczos diagonalization. (a) Corresponds to the PM phase at $U/W = 0.4$, (b) is for the AF2 phase at
$U/W = 4.0$, while (c) is for the AF1 phase at $U/W = 10.0$.}
\label{fig8}
\end{figure}
\begin{figure}[h]
\centering
\rotatebox{0}{\includegraphics*[width=\linewidth]{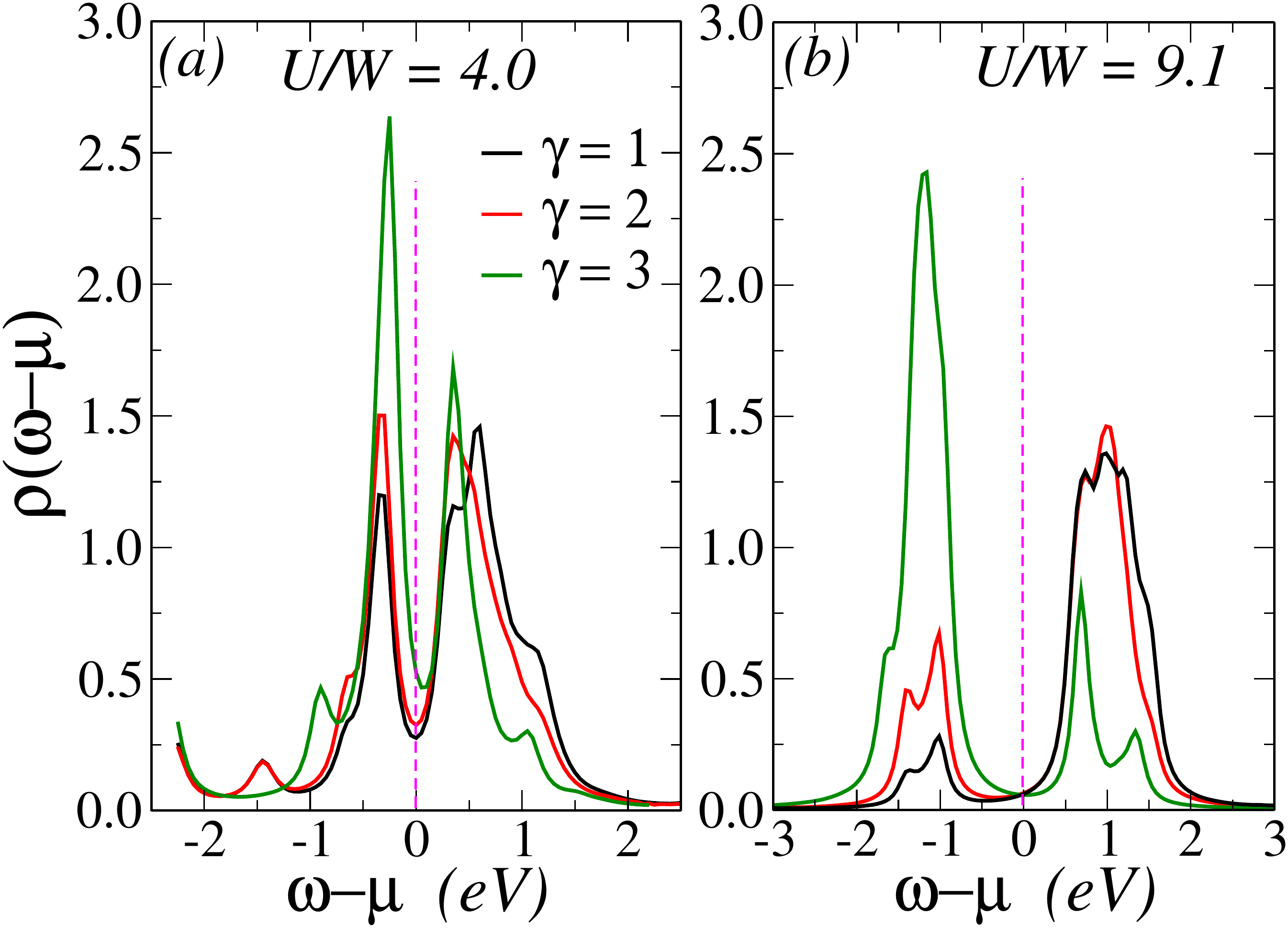}}
\rotatebox{0}{\includegraphics*[width=\linewidth]{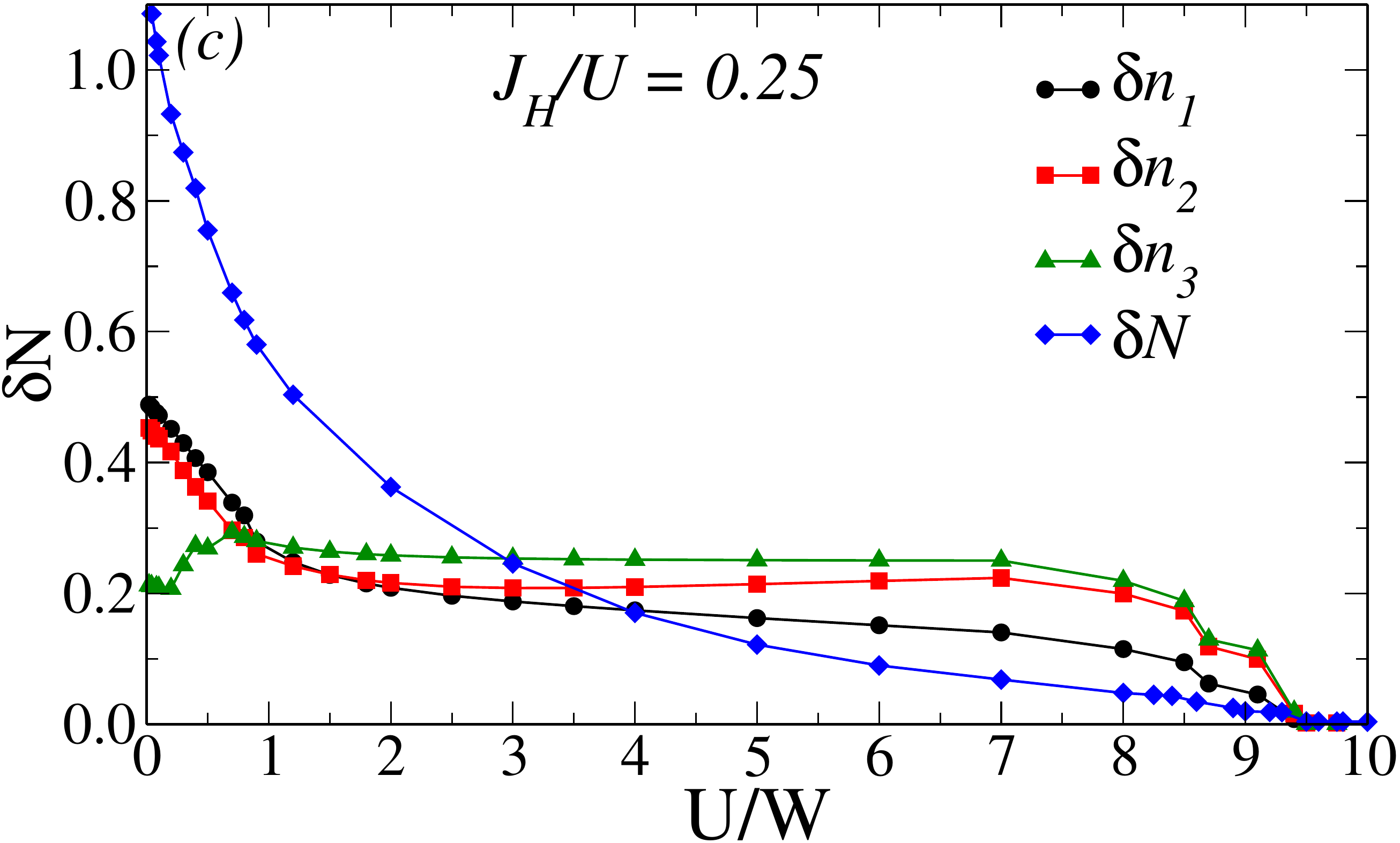}}
\caption{Local DOS of different orbitals corresponding to different phases at $J_H/U = 0.25$ for system size $L=16$,
using dynamical-DMRG. (a) is for the AF2 phase at $U/W=4.0$ and (b) for the AF3 phase at $U/W=9.1$.
(c) Site-averaged local charge fluctuations and  orbital-resolved charge fluctuations vs  $U/W$,  at $J_H/U = 0.25$
and for $L=16$. The nonzero values indicate charge fluctuations are present in the entire AF2 phase, suggesting it is metallic.}
\label{fig7}
\end{figure}

To characterize, at least qualitatively, the metallic vs. insulating nature of the 
different phases, we have calculated the orbital-resolved density-of-states, using the Lanczos method for small 
$L=4$ three-orbital Hubbard model clusters. While these clusters are small, the results are exact. Figure~\ref{fig8}  
contains the orbital-resolved density-of-states (DOS)
vs  $\omega-\mu$ ($\omega$ is the frequency and $\mu$ the chemical potential),
 for different values of the interaction parameter $U/W$. In the paramagnetic phase,
all the three orbitals have a robust weight at the Fermi level, Fig.~\ref{fig8}(a), indicating metallic behavior. 
For the block phase at $U/W=4$, we observe considerably lower weight at the Fermi level for all the three orbitals, 
Fig.~\ref{fig8}(b), signaling a possible pseudogap and bad metallic behavior in the system. 
As expected, in the Mott phase Fig.~\ref{fig8}(c) shows that at $U/W=10$ the system opens a large gap, 
confirming the insulating nature of the AF1 state. The lower Hubbard band of insulating orbitals 1 and 2 is 
not shown (located much lower in energy).

To understand better the characteristics of metallic vs. insulating behaviour, in addition to Lanczos 
we have calculated the  orbital-resolved local density of state $\rho_{i,\gamma}(\omega)$ as a function of frequency $\omega$
using dynamical-DMRG within the correction-vector formalism in Krylov space~\cite{nocera1}.
The orbital-resolved local-density of state (LDOS) has two components: (i) Above the chemical potential it becomes
\begin{equation}
\rho^{+}_{i,\gamma}(\omega)=\frac{-1}{\pi} Im\left[ \left< \psi_0 \left| c_{i,\gamma} \frac{1}{\omega -H +E_g+i\eta}c_{i,\gamma}^{\dagger} \right|\psi_0 \right> \right],
\end{equation}
and (ii) below the chemical potential the LDOS is
\begin{equation}
\rho^{-}_{i,\gamma}(\omega)=\frac{1}{\pi} Im\left[ \left< \psi_0 \left| c_{i,\gamma}^{\dagger} \frac{1}{\omega +H -E_g-i\eta}c_{i,\gamma} \right|\psi_0 \right> \right],
\end{equation}
where $c_{i,\gamma}$ is the fermionic anihilation operator while $c^{\dagger}_{i,\gamma}$ is the creation operator, 
$E_g$ is the ground state energy, and $\psi_0$ is the ground-state wave function of the system. We set the broadening parameter as $\eta=0.1$
for the DDMRG calculations. To avoid edge effects, for the LDOS we chose a central site $i=L/2+1$ for the system size $L=16$. 
For the block phase at $U/W=4.0$, [Fig.~\ref{fig7}(a)], a pseudogap with supressed weight near the Fermi-energy appears,
which is in accord with the Lanczos DOS, suggesting a bad metalic behavior for the AF2 phase.
In Fig.~\ref{fig7}(b), results for the LDOS at the AF3 phase using $U/W=9.1$ are shown.
Here, due to the appearance of orbital-selective density order, we calculate results for two sites (one of each kind, i.e. with $\gamma=3$
equal to 2.0 and 1.5) and then
average to obtain a net LDOS. The resulting LDOS at $U/W=9.1$ in Fig.~\ref{fig7}(b)  indicates insulating behaviour of the system. 

In addition to the DOS we have also investigated the charge fluctuations $\delta N$, to distinguish between a metal and an insulator.
Figure~\ref{fig7}(c) displays the $\delta N$ charge fluctuations defined as $\delta N= 1/L \sum_i\left(\langle n^2_i \rangle -\langle n_i \rangle^2 \right)$ 
(where $n_i=\sum_{\gamma} n_{i \gamma}$) and also the orbital-resolved charge fluctuation 
$\delta N_{\gamma}= \frac{1}{L}\sum_i\langle n^2_{\gamma,i}\rangle - \langle n_{\gamma,i}\rangle^2$ varying $U/W$.
For $U/W \lesssim 0.8$, the large local charge fluctuations indicate strong metallic behavior 
in the PM phase as expected. Increasing $U/W$, the charge fluctuations $\delta N$ decrease substantially but remain finite 
for $U/W \lesssim 8.5$, hinting towards a (bad) metallic behavior of the system in the block AF2-phase. 
Moving beyond $U/W>9.5$, the charge fluctuations approach zero, providing further evidence of insulating behavior
in the AF1-phase. The AF3 phase is difficult to judge because of its narrow range nature, but it also seems insulating. 
These results are in agreement with the Lanczos DOS analysis in Fig.~\ref{fig8}.

\subsection{Phase Diagram varying  $J_H/U$ and $U/W$}
\begin{figure}[h]
\centering
\rotatebox{0}{\includegraphics*[width=\linewidth]{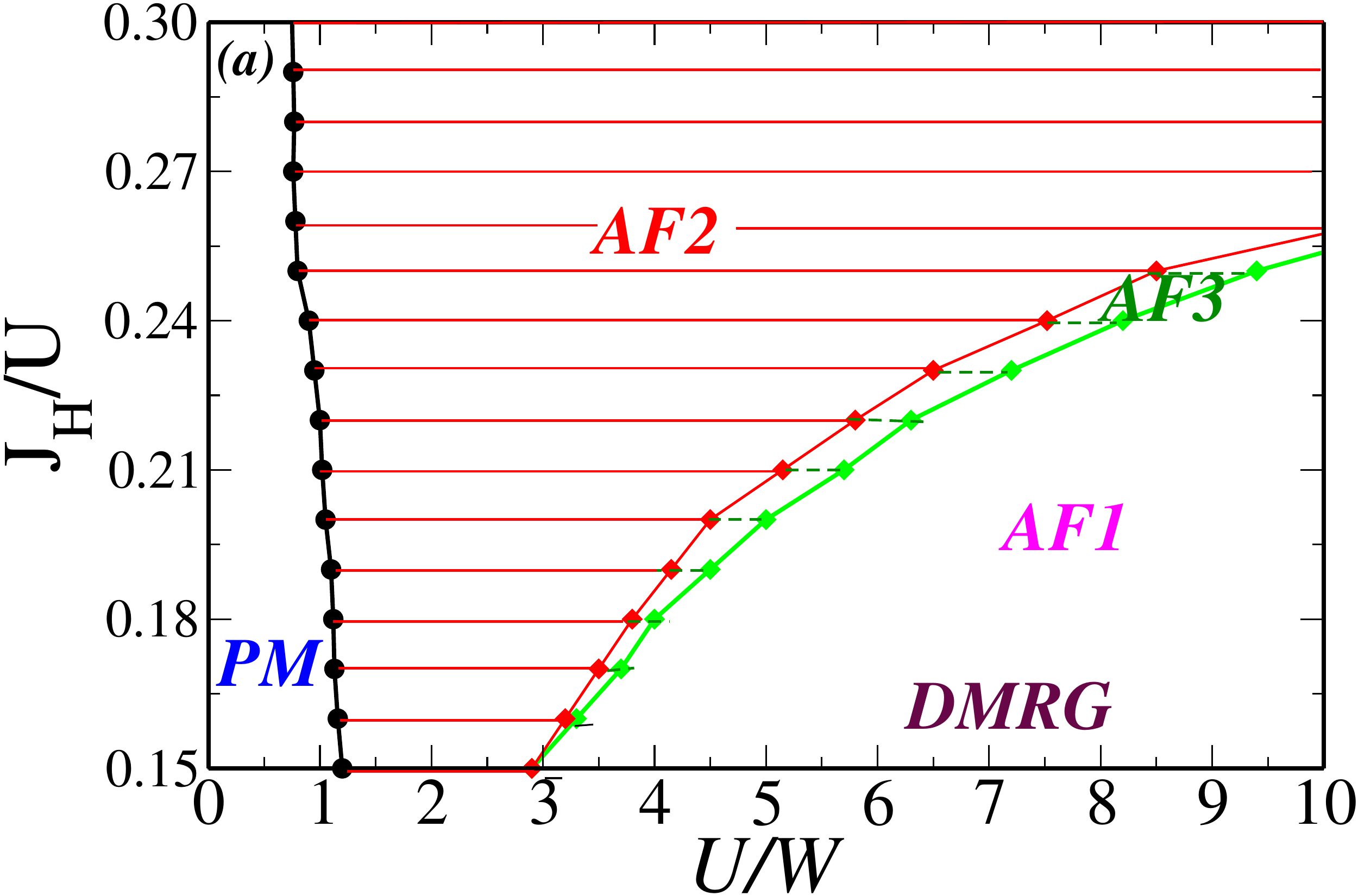}}
\rotatebox{0}{\includegraphics*[width=\linewidth]{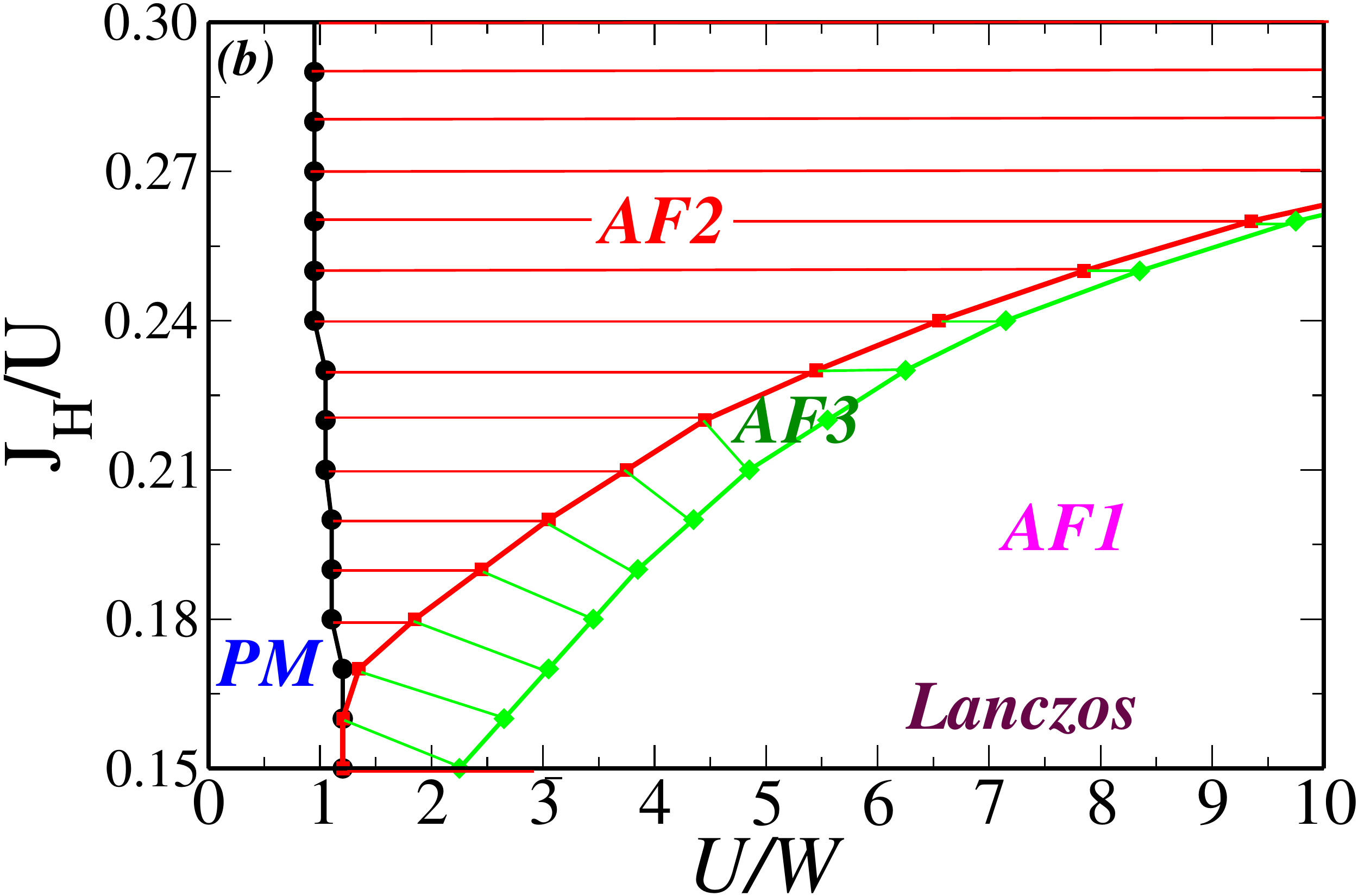}}
\caption{Phase diagram of the three-orbital Hubbard model with the hopping amplitudes of Na$_2$FeSe$_2$,
varying the Hund coupling and Hubbard interactions. Panel (a) depicts results based on DMRG while panel 
(b) are results using the Lanczos method on a $L=4$ sites cluster and open boundary conditions. 
PM stands for paramagnetic phase, while AF2 for block phase with 
$\uparrow \uparrow \downarrow \downarrow$ order. 
The intermediate phase AF3 with $\uparrow \uparrow \downarrow \uparrow \downarrow \downarrow \uparrow \downarrow$
spin ordering appears using both methods in a narrow range of couplings. AF1 stands for the 
staggered antiferromagnetic phase $\uparrow \downarrow \uparrow \downarrow  \uparrow$.}
\label{fig9}
\end{figure}

 Figure~\ref{fig9} contains the  phase diagrams of our three-orbital Hubbard 
model using realistic hopping parameters for  Na$_2$FeSe$_2$ and varying $J_H/U$ from 0.15 to 0.30 and $U/W$ from 0 
to 10. The phase diagram shown in Fig.~\ref{fig9}(a) is based on the DMRG calculations ($L=16$),
 while Fig.~\ref{fig9}(b) is based on Lanczos calculations using $L=4$ sites. 
To obtain the phase boundaries among the different phases, we have used the peak values 
of the spin-structure factor $S(q_{p})$ and the site-average occupancies of each of the orbitals
$n_{\gamma}= \frac{1}{L}\sum_{i,\sigma} \langle n_{i\sigma \gamma} \rangle$.
For lower values of  $U/W$, as expected the metallic PM  phase dominates in the phase diagram for any values
of $J_H/U$. The phase boundary of the PM phase
clearly is very similar between the DMRG and Lanczos results. 
Further increasing the Hubbard interaction $U/W$, in the lower range of Hund couplings $J_H/U$ shown,
the block AF2 phase stabilizes in a small region of the phase diagram,
while the staggered AF1 phase dominates over a larger portion. At not too large $J_H/U$, the superexchange 
mechanism dominates and promotes primarily staggered AF1 magnetic ordering, as expected. 
For these moderate values of $J_H/U$, a rapid cascade of transitions (PM $\to$ AF2 $\to$ AF3 $\to$ AF1) is observed.
For $J_H/U<0.19$, the narrow region in between AF2 $\to$ AF1
shows incommensurate behavior (not shown), while for $J_H/U>0.19$ this intermediate region
displays the exotic AF3-spin order with  peak at $q=3\pi/4$. 

Interestingly, by increasing $J_H/U$ the block AF2 phase with spin configuration $\uparrow \uparrow \downarrow \downarrow$
stabilizes over a large portion of the phase diagram. This magnetic block state (AF2) is the same as 
found before in the context of orbital-selective Mott phases~\cite{julian,herbrychNatComm,herbrychblock}, although here the three orbitals 
remain itinerant, i.e. none has a population locked to one. As in those previous efforst,
we believe the block spin order AF2 arises from competing superexchange order at small $J_H$
and double-exchange ferromagnetism at large $J_H$. While in our phase diagram 
there is no ferromagnetic phase in the range studied, we found that removing the NNN hopping leads to a stable
ferromagnetic region, as in previous efforts~\cite{julian,herbrychNatComm,herbrychblock}. Thus, the ferromagnetic state is certainly close in energy.

\begin{figure}[h]
\centering
\rotatebox{0}{\includegraphics*[width=\linewidth]{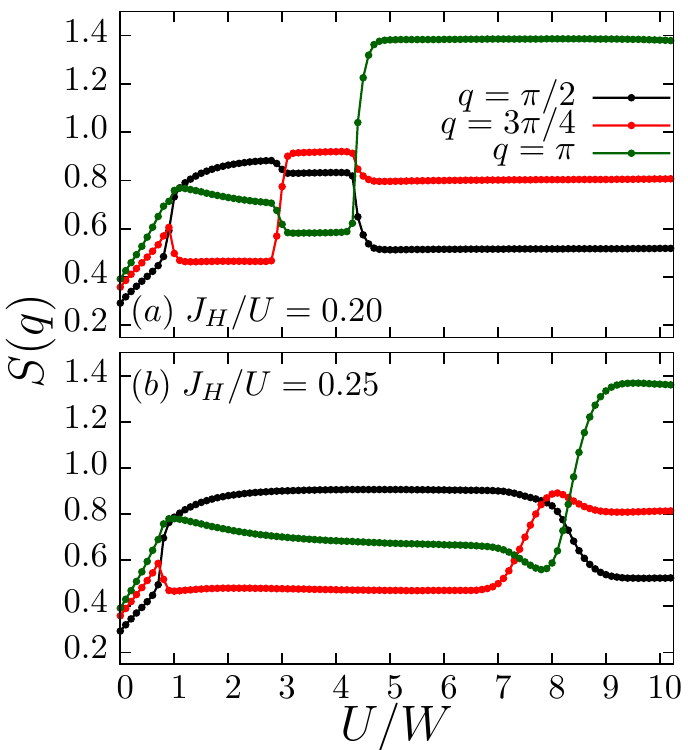}}
\caption{Spin structure factors $S(q)$ at $q=\pi/2$, $3\pi/4$, and $\pi$ vs. $U/W$ 
for (a) $J_H/U=0.20$ and (b) $J_H/U=0.25$, using the Lanczos method for $L=4$ sites, and the 
three-orbitals Hubbard model used here.} 
\label{fig10}
\end{figure}

Also note the good agreement between the  DMRG and Lanczos results found
for the phase diagrams, see Fig.~\ref{fig9}(a) vs Fig.~\ref{fig9}(b), except for small $J_H/U$ where 
the AF3 phase is broader with Lanczos than DMRG, with opposite effects for the AF2 region. 
This small difference may be due to size effects.
However, at moderate $J_H/U$ between 0.19 and 0.25 -- a region considered realistic for iron-based compounds -- 
the AF2 phase, which represents our main prediction for the physics of Na$_2$FeSe$_2$ if ever synthesized, is large and robust
as Fig.~\ref{fig4}(a) shows using DMRG and Fig.~\ref{fig10}(b) using Lanczos.

\section{V. Conclusions}
In this publication, the phase diagram of the one-dimensional chain  compound Na$_2$FeSe$_2$ has been investigated.
We used a realistic three-orbital Hubbard model with the hopping amplitudes derived from {\it ab initio} calculations.
The phase diagram presented here was constructed at electronic density $n=4$ per site (the analog of $n=6$ in a five-orbital system).
To our best knowledge, this is the first material of the family of iron superconductors that has both a dominant
chain geometry in the structure (not ladder) and valence Fe$^{2+}$. Our phase diagram is based primarily on  DMRG measurements of the orbital 
occupancy and  spin structure factor, supplemented by Lanczos techniques.
In comparison to previously studied $n=5$ one-dimensional three-orbital models  for iron based compounds such as TlFeSe$_2$, 
which display a trivial staggered spin order, we find a much richer phase diagram for the  
alkali metal iron selenide compound Na$_2$FeSe$_2$. 
In particular, at low $J_H/U$ the staggered spin order dominates, but 
increasing $J_H/U$ the block AF2 phase $\uparrow \uparrow \downarrow \downarrow$ 
is stabilized over a large region of the phase diagram. 
We also observed a narrow region of a new phase AF3, with charge density wave properties and a combination of
features of the AF1 and AF2 dominant phases, leading to a net $\uparrow \uparrow \downarrow \uparrow \downarrow \downarrow \uparrow \downarrow$ magnetic order. Previous results with iron ladders suggest that high pressure probes may also bring surprises,
such as metallicity and even superconductivity. As a consequence, we encourage experimentalists to synthesize Na$_2$FeSe$_2$ and
investigate its magnetic properties via neutron scattering experiments.

\section{Acknowledgments}
We thank Yang Zhang for useful discussions.
The work of B.P, R.S, N.K., and E.D. was supported by the U.S. Department of
Energy (DOE), Office of Science, Basic Energy Sciences
(BES), Materials Sciences and Engineering Division. L.-F.L. was supported by the National Natural 
Science Foundation of China (Grants No. 11834002 and No.11674055) and by the China Scholarship Council.
G.A. was partially supported by the Center for Nanophase Materials Sciences, which is a U.S. DOE Office of Science User Facility, and by the Scientific Discovery  through  Advanced  Computing  (SciDAC) program  funded  by  U.S.  DOE,  Office  of  Science, Advanced  Scientific  Computing  Research  and  BasicEnergy Sciences, Division of Materials Sciences and Engineering. J.H. acknowledges grant support by the Polish National Agency of Academic Exchange (NAWA)
under  contract  PPN/PPO/2018/1/00035. Validation and some computer runs were conducted at the Center for Nanophase Materials Sciences, which is a DOE Office of Science User Facility.

\end{document}